\documentclass[aps,prx,twocolumn,10pt,groupedaddress]{revtex4-1}
\usepackage{amssymb}
\usepackage{graphicx}
\usepackage{amsmath}
\usepackage{hyperref}
\usepackage{subfigure}
\usepackage{float}
\usepackage{color}
\usepackage{ulem}
\usepackage[utf8]{inputenc}

\begin{document}

\title{Chameleon dark energy can resolve the Hubble tension}
\author{Rong-Gen Cai$^{1,2}$}
\email{cairg@itp.ac.cn}
\author{Zong-Kuan Guo$^{1,2}$}
\email{guozk@itp.ac.cn}
\author{Li Li$^{1,2}$}
\email{liliphy@itp.ac.cn (corresponding author)}
\author{Shao-Jiang Wang$^{1}$}
\email{schwang@itp.ac.cn (corresponding author)}
\author{Wang-Wei Yu$^{1,3}$}
\email{yuwangwei@mail.itp.ac.cn}
\affiliation{$^1$CAS Key Laboratory of Theoretical Physics, Institute of Theoretical Physics, Chinese Academy of Sciences, P.O. Box 2735, Beijing 100190, China}
\affiliation{$^2$School of Fundamental Physics and Mathematical Sciences, Hangzhou Institute for Advanced Study (HIAS), University of Chinese Academy of Sciences, Hangzhou 310024, China}
\affiliation{$^3$School of Physical Sciences, University of Chinese Academy of Sciences (UCAS), Beijing 100049, China}

\begin{abstract}
Values of the Hubble constant between the direct measurements from various independent local observations and that inferred from the cosmic microwave background with the $\Lambda$-cold-dark-matter model are in tension with persistent significance.
We propose a late-time inhomogeneous resolution suggesting that a chameleon field coupled to a local overdensity of matter could be trapped at a higher potential energy density as an effective cosmological constant driving the local expansion rate faster than that of the background with lower matter density. We illustrate this mechanism in a toy model in which a region with only $20\%$ overdensity of matter is sufficient to resolve the Hubble tension, and the Hubble constant measured by the local distance ladders could be accommodated by the chameleon coupled to the observed overdensities from the large-scale structure surveys.
\end{abstract}
\maketitle

\section{Introduction} 

The Hubble tension \cite{Bernal:2016gxb,Verde:2019ivm,Knox:2019rjx,Riess:2020sih,DiValentino:2020zio} is now becoming a pressing issue for the $\Lambda$-cold-dark-matter ($\Lambda$CDM) model reconciling the model-inferred value of the Hubble constant $H_0=67.27\pm0.60\,\mathrm{km\,s^{-1}\,Mpc^{-1}}$ from the cosmic microwave background (CMB) \cite{Henning:2017nuy,Aghanim:2018eyx,Aiola:2020azj} as well as extragalactic $\gamma$-ray background \cite{Dominguez:2019jqc} measurements with respect to various independent local direct measurements either from supernovae (SNe) Ia data (see also \cite{deJaeger:2020zpb} for type II SNe) calibrated by Cepheid \cite{Riess:2016jrr,Riess:2018byc,Riess:2018uxu,Riess:2019cxk,Riess:2020fzl} ($H_0=73.2\pm1.3\,\mathrm{km\,s^{-1}\,Mpc^{-1}}$ from most recent measurement \cite{Riess:2020fzl}), tip of the red giant branch (TRGB) \cite{Freedman:2019jwv,Yuan:2019npk,Freedman:2020dne,Soltis:2020gpl}, and Miras \cite{Huang:2019yhh}
or from masers \cite{Huang:2019yhh,Pesce:2020xfe}, surface brightness fluctuations (SBFs) \cite{Khetan:2020hmh,Blakeslee:2021rqi}, the baryonic Tully–Fisher relation \cite{Kourkchi:2020iyz,Schombert:2020pxm}, parallax measurement of quasar 3C 273 \cite{Wang:2019gaq},  gravitational-wave standard sirens \cite{Abbott:2017xzu,Abbott:2019yzh,Mukherjee:2019qmm,Wang:2020vgr}, and strong lensing time delay (SLTD) \cite{Wong:2019kwg,Shajib:2019toy,Birrer:2020tax} that are largely independent of the local distance ladder calibrations. 
This Hubble tension is also manifested in the dubbed inverse distance ladder, where a degeneracy of $r_\mathrm{d}H_0$ \cite{Heavens:2014rja,Cuesta:2014asa,Verde:2016ccp} measured by the SNe+baryon acoustic oscillation (BAO) could be broken by a CMB-inferred prior on the comoving sound horizon $r_\mathrm{d}$ at drag epoch \cite{Vonlanthen:2010cd,Audren:2013nwa,Audren:2012wb,Cuesta:2014asa,Aubourg:2014yra,Verde:2016wmz,Alam:2016hwk,Macaulay:2018fxi,Alam:2020sor} (see also \cite{Pogosian:2020ded} for methods of breaking the $r_\mathrm{d}H_0$ degeneracy from $\Omega_\mathrm{m}h^2$ priors), leading to a lower $H_0=67.8\pm1.3\,\mathrm{km\,s^{-1}\,Mpc^{-1}}$ from SNe+Dark Energy Survey \cite{Macaulay:2018fxi} that is consistent with the CMB-independent constraint on $H_0$ from BAO+big bang nucleosynthesis (BBN)  \cite{Addison:2013haa,Aubourg:2014yra,Addison:2017fdm,Blomqvist:2019rah,Cuceu:2019for,Schoneberg:2019wmt,Philcox:2020vvt,DAmico:2020kxu}. Furthermore, the CMB-inferred prior on $r_\mathrm{d}$ is also in tension with the joint BAO+SNe+$H_0$ distance ladder constraint \cite{Bernal:2016gxb,Knox:2019rjx}. See also \cite{Lin:2019htv} for an alternative view of the Hubble tension in the $H_0-\Omega_\mathrm{m}$ plane.

The current status of observations has resulted in a dilemma in which the early-time global-fitting constraints on $H_0$ are robust with or without CMB data, and there seems to be no single systematic to incorporate all late-time local direct measurements on the Hubble constant since they are largely independent. 
However, there are still some noticeable features that might guide us along the right path.
The first feature subjected to the early-time measurement is the offset of the CMB constraint on $H_0$ from the low-$\ell$ and high-$\ell$ data \cite{Aylor:2017haa,Henning:2017nuy,Knox:2019rjx} as well as a  relatively high $H_0$ value from CMB $E$-mode data alone \cite{Dutcher:2021vtw,Addison:2021amj}, suggesting possible new physics at small scales that might be relevant to the Hubble tension.
The second feature subjected to the late-time measurements is the quasilocal measurements along the Hubble flow on the Hubble parameter from cosmic chronometers \cite{Chen:2016uno,Farooq:2016zwm,Yu:2017iju}, with a preference for a smaller Hubble constant close to the CMB-inferred value.
The third one subjected to the local measurements is the reduced Hubble constant $H_0=67.4_{-3.2}^{+4.1}\,\mathrm{km\,s^{-1}\,Mpc^{-1}}$ from the improved  joint hierarchical analysis of the TDCOSMO+SLACS sample for SLTD \cite{Birrer:2020tax} with respect to the H0LiCOW result \cite{Wong:2019kwg} $H_0=73.3_{-1.8}^{+1.7}\,\mathrm{km\,s^{-1}\,Mpc^{-1}}$. It seems that a smaller Hubble constant is preferably measured when one uses more distant indicators out to $\gtrsim\mathcal{O}(1)$ Gpc.

On the other hand, the proposed solutions to the Hubble tension (see \cite{DiValentino:2020zio,DiValentino:2021izs} and references therein for a list of these proposals) has also resulted in a dilemma that in which all the early-time solutions given by solely reducing the cosmic sound horizon cannot fully resolve the Hubble tension \cite{Jedamzik:2020zmd} (see, however, \cite{Bernal:2020vbb}) without conflicting with the galaxy clustering data \cite{Alam:2016hwk} or galaxy weak lensing data \cite{Abbott:2017wau,Asgari:2020wuj} (see also \cite{Seto:2021xua} for BBN constraints), which is in agreement with earlier studies \cite{Hill:2020osr,Ivanov:2020ril,DAmico:2020ods} (see, however, \cite{Smith:2020rxx}) on the early dark energy model \cite{Poulin:2018cxd}. The consistency checks \cite{Philcox:2020xbv,Lin:2021sfs} from the sound horizon constraints among the matter-radiation equality, recombination, and the end of the drag epoch also disfavor early-time solutions. Furthermore, it appears that very little room is allowed for introducing a homogeneous late-time dark energy (except for the interacting dark energy model \cite{DiValentino:2019ffd}) to deviate significantly from the $\Lambda$CDM at low redshift \cite{Dhawan:2020xmp}, as it is strongly constrained by the BAO+SNe data \cite{Benevento:2020fev,Efstathiou:2021ocp}. Special care should be taken for two kinds of illusionary solutions: one is the enlarged uncertainties incorporating the Hubble tension rather than a genuine shift in the central value of the  Hubble constant when marginalizing over additional parameters introduced to extend the $\Lambda$CDM \cite{Vagnozzi:2019ezj}, and the other is a naively blind combination of CMB+BAO+SNe with $H_0$ data, without which the proposal hardly improves the fitting of the Hubble constant.

\begin{figure*}
\centering
\includegraphics[width=0.45\textwidth]{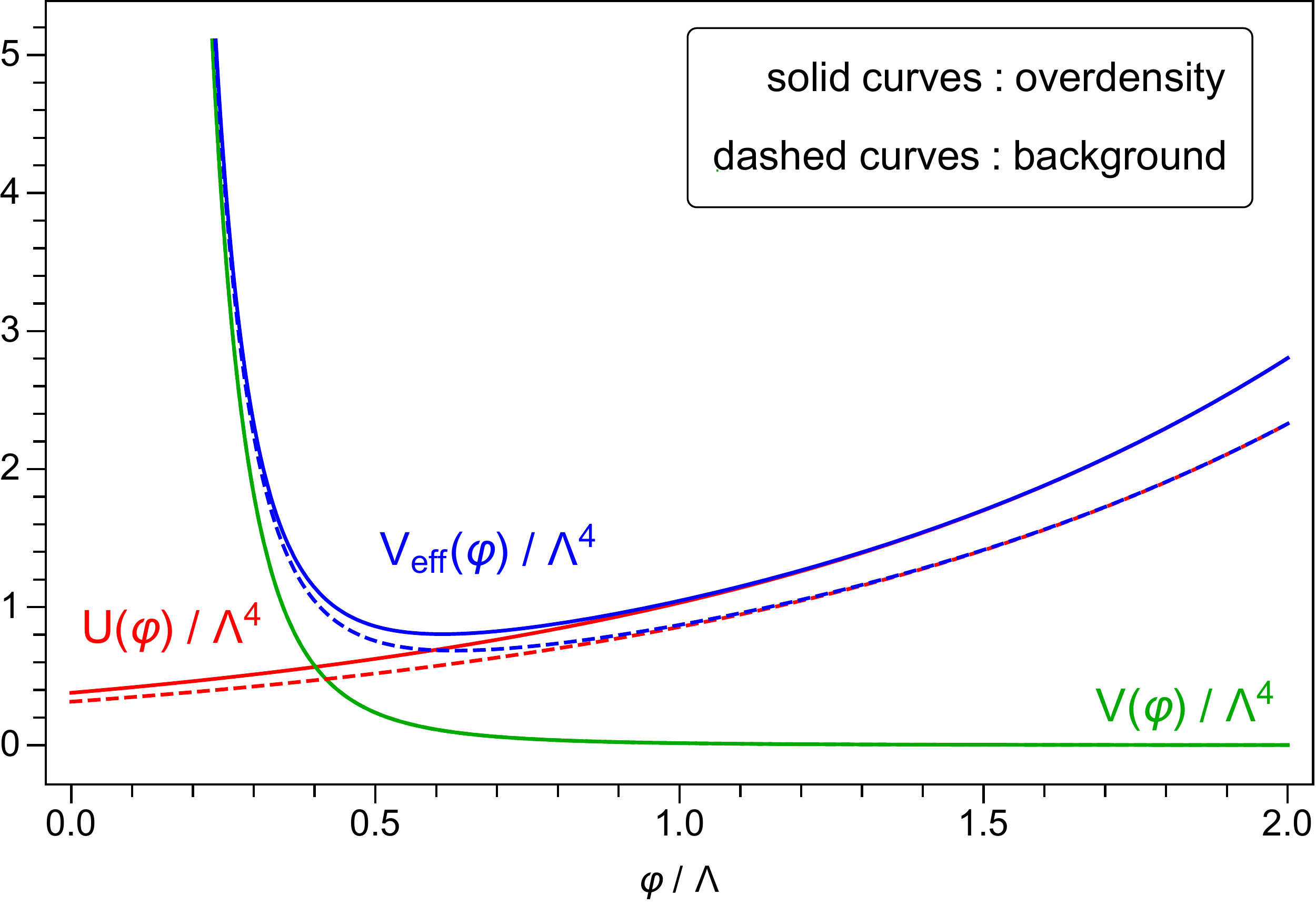}
\includegraphics[width=0.5\textwidth]{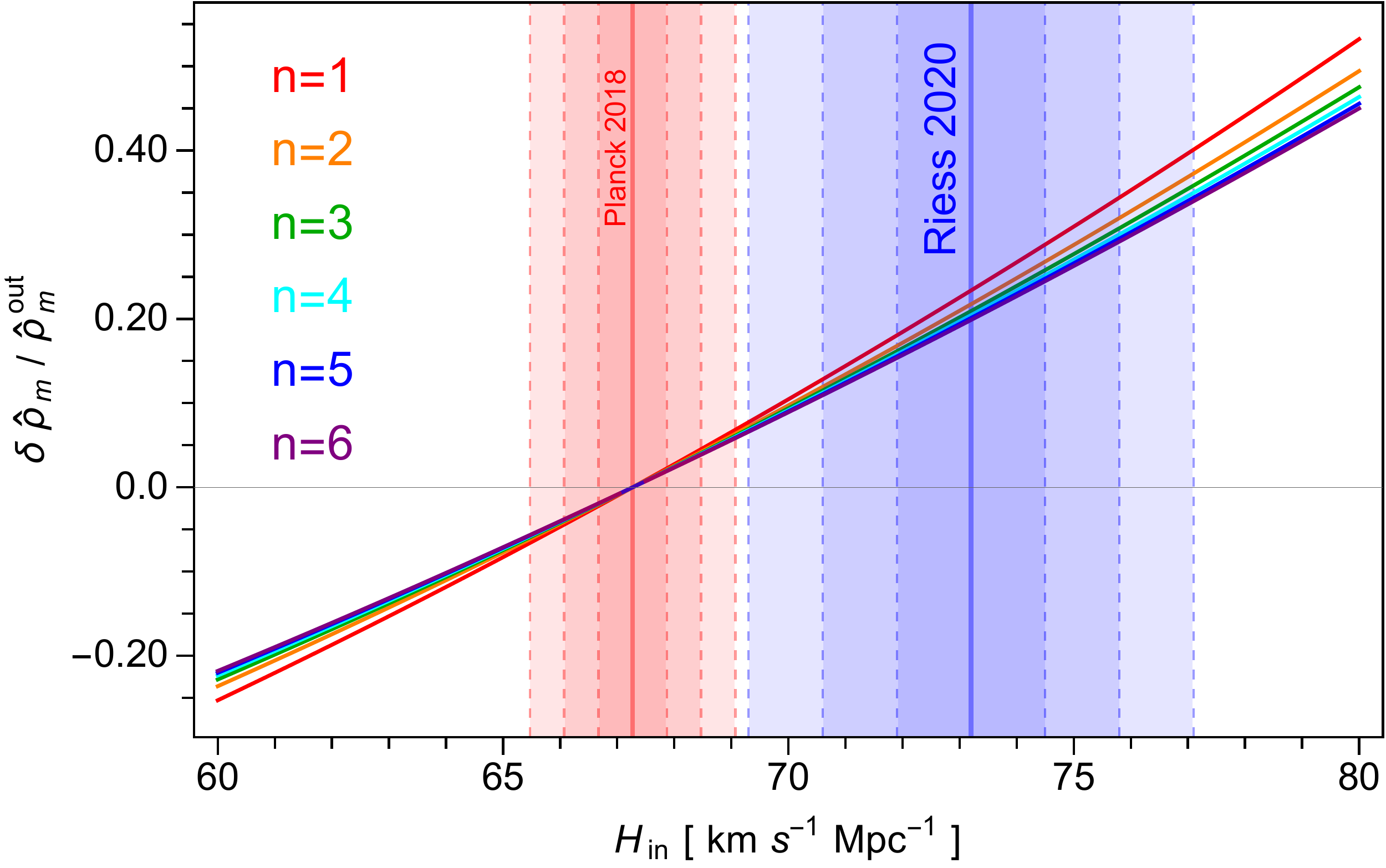}
\caption{Left panel : effective potential (blue curve) of the chameleon field receives larger contributions (red curve) to the original chameleon potential (green curve) from overdensity (solid curves) than underdensity (dashed curves). Right panel :  overdensity of a region with respect to a local Hubble constant for chameleon potentials \eqref{eq:Veff} of different $n$ confronted with the local distance ladder measurement (blue shaded area) and Planck 2018 measurement (red shaded area). }\label{fig:toy}
\end{figure*}

A possible way out of the above dilemma is turning to local inhomogeneous solutions, for example, cosmic voids  \cite{GarciaBellido:2008nz,Keenan:2013mfa,Hoscheit:2018nfl}, which, however, have been strongly constrained by the SNe data \cite{Wojtak:2013gda,Odderskov:2014hqa,Wu:2017fpr,Kenworthy:2019qwq,Lukovic:2019ryg,Cai:2020tpy} (see \cite{Lombriser:2019ahl}, however, for a local underdensity on a 40 Mpc scale that affects the local distance ladder calibrations). We propose here a chameleon field \cite{Khoury:2003aq,Khoury:2003rn,Wang:2012kj,Upadhye:2012vh,Khoury:2013yya} trapped at a higher potential energy density when coupled to an overdensity region of matter could drive the Hubble expansion rate locally larger in that region, as shown in the left panel of Fig. \ref{fig:toy}, from an illustrative toy model given below with a realistic generalization thereafter. It is worth noting that Kenworthy \textit{et al.} \cite{Kenworthy:2019qwq}  imposed a general constraint on any local inhomogeneity, including our model, as we will encounter later. See also \cite{Desmond:2019ygn,Desmond:2020wep} for other chameleon-inspired models of the Hubble tension.

\section{Chameleon dark energy in overdensity}

A toy model of the chameleon field as dark energy is introduced by assuming a runaway potential known as the Peebles-Ratra potential \cite{Peebles:2002gy}
\begin{align}\label{eq:runaway}
V(\varphi)=\alpha\Lambda^4\left(\frac{\Lambda}{\varphi}\right)^n,
\end{align}
and a matter coupling of form
\begin{align}
\Omega(\varphi)=\exp\left(\frac{\varphi}{\Lambda}\right),
\end{align}
so that the effective potential of the chameleon field (see appendix \ref{app:chameleon} or \cite{Cai:2020ndh} for a general derivation), 
\begin{align}\label{eq:Veff}
V_\mathrm{eff}(\varphi)=V(\varphi)+\Omega(\varphi)\hat{\rho}_\mathrm{m},
\end{align}
develops a vacuum-expectation value
\begin{align}
\langle\varphi\rangle=(1+n)\Lambda W\left[\frac{1}{1+n}\left(\frac{\hat{\rho}_\mathrm{m}}{n\alpha\Lambda^4}\right)^{-\frac{1}{1+n}}\right]
\end{align}
to trap the chameleon field at a constant potential energy density so as to play the role of an  effective cosmological constant when $\varphi$ adiabatically tracking the minimum $\langle\varphi\rangle$. Here $W(z)$ is the Lambert function defined by $z=W(z)e^{W(z)}$, and $\alpha$ characterizes the offset of the different mass scales subjected to the field $\varphi$ and potential $V(\varphi)$; namely, any matter coupling of an exponential form
\begin{align}\label{eq:repara1}
\Omega(\varphi)=\exp\left(\beta\frac{\varphi}{\Lambda}\right)\equiv\exp\left(\frac{\varphi}{\bar{\Lambda}}\right)
\end{align}
amounts to the redefinition of $\alpha$ by $\bar{\alpha}\equiv\alpha\beta^{n+4}$ from
\begin{align}\label{eq:repara2}
V(\varphi)=\alpha\beta^{n+4}\bar{\Lambda}^4\left(\frac{\bar{\Lambda}}{\varphi}\right)^n\equiv\bar{\alpha}\bar{\Lambda}^4\left(\frac{\bar{\Lambda}}{\varphi}\right)^n.
\end{align}

For the chameleon field coupled to the background region with the Hubble constant $H_\mathrm{out}=H_\mathrm{CMB}=100\,h_\mathrm{CMB}\,\mathrm{km\,s^{-1}\,Mpc^{-1}}=67.27\,\mathrm{km\,s^{-1}\,Mpc^{-1}}$ inferred from the Planck 2018 measurement \cite{Aghanim:2018eyx}, the Friedmann equation reads
\begin{align}\label{eq:FLRWout1}
3M_\mathrm{Pl}^2H_\mathrm{out}^2=\hat{\rho}_\mathrm{m}^\mathrm{out}+V_\mathrm{eff}(\langle\varphi\rangle_\mathrm{out}; n, \alpha, \hat{\rho}_\mathrm{m}^\mathrm{out}),
\end{align}
where $M_\mathrm{Pl}^{-2}=8\pi G$. The mass scale $\Lambda^4$ will be chosen hereafter as the current critical density $3M_\mathrm{Pl}^2H_\mathrm{out}^2$ for convenience by virtue of the reparametrization freedom from Eqs. \eqref{eq:repara1} and \eqref{eq:repara2} since any redefined $\alpha$ due to the change of $\Lambda$ would eventually be fixed by the boundary condition  \eqref{eq:FLRWout1}, namely,
\begin{align}
1=\Omega_\mathrm{m}^\mathrm{out}+\alpha\left(\frac{\Lambda}{\langle\varphi\rangle_\mathrm{out}}\right)^n+\exp\left(\frac{\langle\varphi\rangle_\mathrm{out}}{\Lambda}\right)\Omega_\mathrm{m}^\mathrm{out},
\end{align}
with 
\begin{align}
\frac{\langle\varphi\rangle_\mathrm{out}}{\Lambda}=(1+n) W\left[\frac{1}{1+n}\left(\frac{\Omega_\mathrm{m}^\mathrm{out}}{n\alpha}\right)^{-\frac{1}{1+n}}\right],
\end{align}
where $\alpha$ could be directly solved for a given $n$ and $\Omega_\mathrm{m}^\mathrm{out}\equiv(\Omega_bh_\mathrm{CMB}^2+\Omega_ch_\mathrm{CMB}^2)/h_\mathrm{CMB}^2=0.315$ inferred from the Planck 2018 measurements \cite{Aghanim:2018eyx} on the baryonic matter $\Omega_bh^2=0.02236$ and cold dark matter $\Omega_bh^2=0.1202$. 

On the other hand, for a chameleon field coupled to a sufficiently large region of the matter overdensity so that the Friedmann equation could be applied,
\begin{align}\label{eq:FLRWin1}
3M_\mathrm{Pl}^2H_\mathrm{in}^2=\hat{\rho}_\mathrm{m}^\mathrm{in}+V_\mathrm{eff}(\langle\varphi\rangle_\mathrm{in}; n, \alpha, \hat{\rho}_\mathrm{m}^\mathrm{in}),
\end{align}
namely,
\begin{align}\label{eq:FLRWin2}
\frac{H_\mathrm{in}^2}{H_\mathrm{out}^2}=\Omega_\mathrm{m}^\mathrm{in}+\alpha\left(\frac{\Lambda}{\langle\varphi\rangle_\mathrm{in}}\right)^n+\exp\left(\frac{\langle\varphi\rangle_\mathrm{in}}{\Lambda}\right)\Omega_\mathrm{m}^\mathrm{in},
\end{align}
with
\begin{align}\label{eq:vevin}
\frac{\langle\varphi\rangle_\mathrm{in}}{\Lambda}=(1+n) W\left[\frac{1}{1+n}\left(\frac{\Omega_\mathrm{m}^\mathrm{in}}{n\alpha}\right)^{-\frac{1}{1+n}}\right].
\end{align}
The overdensity fraction $\Omega_\mathrm{m}^\mathrm{in}\equiv\hat{\rho}_\mathrm{m}^\mathrm{in}/(3M_\mathrm{Pl}^2H_\mathrm{out}^2)\equiv\hat{\rho}_\mathrm{m}^\mathrm{in}/\Lambda^4$ could be solved for given $n$, $H_\mathrm{in}$, and the presolved $\alpha(n, \Omega_\mathrm{m}^\mathrm{out})$. Therefore, the required overdensity 
\begin{align}
\frac{\delta\hat{\rho}_\mathrm{m}}{\hat{\rho}_\mathrm{m}^\mathrm{out}}\equiv\frac{\Omega_\mathrm{m}^\mathrm{in}-\Omega_\mathrm{m}^\mathrm{out}}{\Omega_\mathrm{m}^\mathrm{out}}
\end{align}
is obtained for given $n$, $\Omega_\mathrm{m}^\mathrm{out}$, and $H_\mathrm{in}$ that matches the local measurements. As a benchmark estimation, for $n=4$ and $\Omega_\mathrm{m}^\mathrm{out}=0.315$, a local region with overdensity $\delta\hat{\rho}_\mathrm{m}/\hat{\rho}_\mathrm{m}^\mathrm{out}=20\%$ is sufficient to reproduce the local measurement on the Hubble constant $H_\mathrm{in}=73.20\,\mathrm{km\,s^{-1}\,Mpc^{-1}}$ \cite{Riess:2020fzl}. In general, we plot the local overdensity with respect to the local Hubble constant in the right panel of Fig. \ref{fig:toy}, which is robust to the choices of different runaway potentials with $n=1, 2, 3, 4, 5, 6$. A simple fitting formula reads
\begin{align}
\frac{\delta\hat{\rho}_\mathrm{m}}{\hat{\rho}_\mathrm{m}^\mathrm{out}}
&=\left(0.65e^{-0.57n}+2.13\right)\left(\frac{H_\mathrm{in}}{67.27}-1\right)\nonumber\\
&+\left(0.65e^{-0.27n}+1.07\right)\left(\frac{H_\mathrm{in}}{67.27}-1\right)^2.
\end{align}
Since $H_\mathrm{in}$ and $\Omega_\mathrm{m}^\mathrm{in}$ could be inferred from one another for given $\Omega_\mathrm{m}^\mathrm{out}\equiv\Omega_\mathrm{m}^\mathrm{CMB}=0.315$ and is insensitive to the choice of $n$, our chameleon dark energy model is therefore effectively a zero-parameter model if we further fix, for example, $n=4$ hereafter. Other chameleon potential could also work as long as it is monotonically decreasing.

\section{Chameleon dark energy in our local Universe}

\begin{figure*}
\centering
\includegraphics[width=0.53\textwidth]{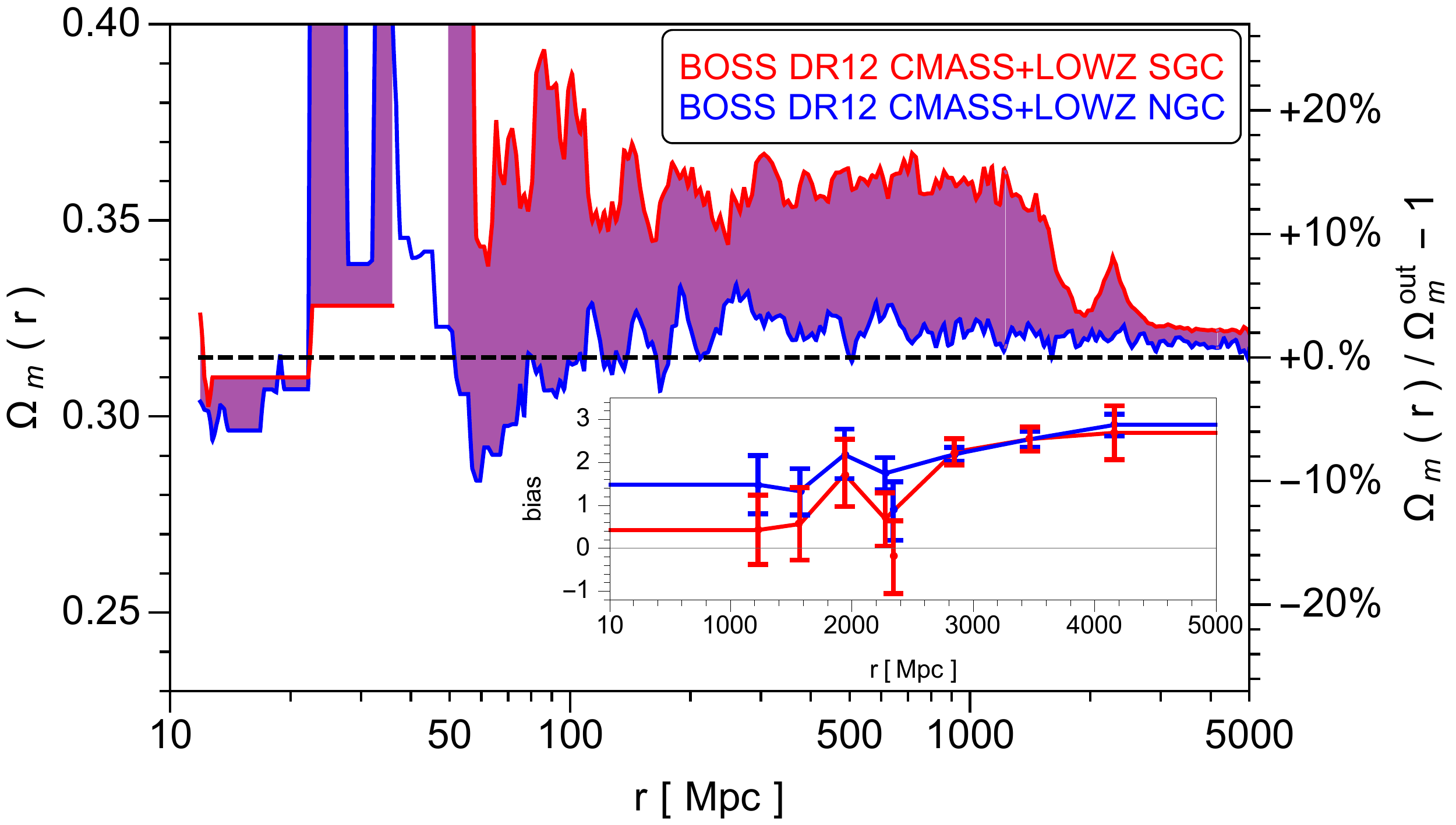}
\includegraphics[width=0.46\textwidth]{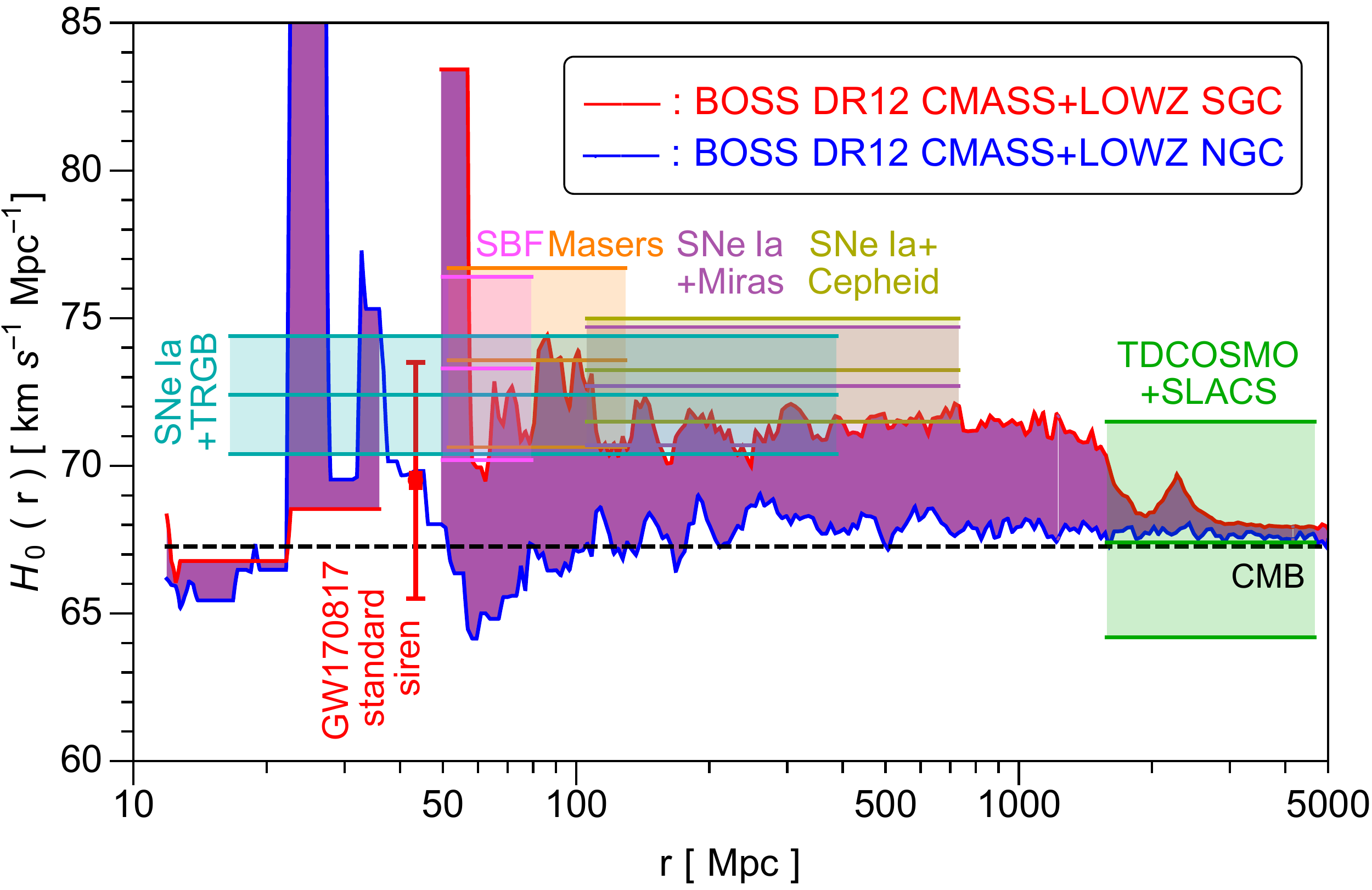}\\
\caption{Left panel: local profile of the matter density fraction from BOSS DR12 LOWZ+CMASS NGC (blue) and SGC (red) samples. Inset:  galaxy bias borrowed from \cite{Lavaux:2019fjr}. Right panel: inferred value of the Hubble constant from the BOSS DR12 LOWZ+CMASS NGC (blue) and SGC (red) samples with respect to the various measurements, as indicated.}\label{fig:LocalUniverse}
\end{figure*}

We next turn to a more realistic setup for our local Universe described by an inhomogeneous but isotropic Lema\^{i}tre-Tolman-Bondi (LTB) metric \cite{Lemaitre:1933gd,Tolman:1934za,Bondi:1947fta}. The resulted inhomogeneous profile for the local Hubble constant could be solved from the LTB version of the Friedmann equation \eqref{eq:FLRWin1} at the present time (see appendix \ref{app:LTB} for a derivation) assuming the spatial flatness for simplicity, 
\begin{align}\label{eq:FRW0}
\frac{H_0(r)^2}{H_\infty^2}=\Omega_m(r)+\alpha\left(\frac{\Lambda}{\langle\varphi\rangle(r)}\right)^{n}+\exp\left(\frac{\langle\varphi\rangle(r)}{\Lambda}\right)\Omega_\mathrm{m}(r),
\end{align}
where $H_0(r)\equiv H(t_0,r)$, $H_\infty\equiv H(t_0,r\to\infty)\equiv H_\mathrm{CMB}$ at the present time $t_0$ and
\begin{align}
\frac{\langle\varphi\rangle(r)}{\Lambda}=(1+n)W\left[\frac{1}{1+n}\left(\frac{\Omega_\mathrm{m}(r)}{n\alpha}\right)^{-\frac{1}{1+n}}\right],
\end{align}
with $\alpha$ solved from matching $H_0(r)$ to $H_\mathrm{CMB}$ at $r\to\infty$ first. This also justifies the depiction we used for Eq. \eqref{eq:FLRWin2} and the right panel of Fig. \ref{fig:toy} that follows. To estimate the local matter density fraction profile
\begin{align}
\Omega_\mathrm{m}(r)=(1+\delta_\mathrm{m}(r))\frac{\hat{\rho}_\mathrm{m}(\infty)}{3M_\mathrm{Pl}^2H_\infty^2}\equiv(1+\delta_\mathrm{m}(r))\Omega_\mathrm{m}^\mathrm{CMB}
\end{align}
from the matter density contrast profile $\delta_\mathrm{m}(r)$, we use the galactic tracer for the matter distribution via a galaxy bias parameter $b$ \cite{Kaiser:1984sw} by
\begin{align}\label{eq:bias}
\delta_\mathrm{m}(r)=\frac{\delta_\mathrm{g}(r)}{b}.
\end{align}

The galaxy density contrast profile $\delta_g(r)$ is approximated here by the number density contrast profile from naively counting the number excess of data galaxies with respect to the random galaxies around us,
\begin{align}
\delta_g(r)=\frac{N_D(r)}{N_R(r)}-1,
\end{align}
where the number countings $N_D(r)$ and $N_R(r)$ for data and random galaxies, respectively, are weighted by
\begin{align}
N_G(r)=\sum\limits_{G_i}w(G_i)\Theta_r(d(G_i)),\quad G=D,R,
\end{align}
with the distance to galaxies $d(G_i)$ averaged in a spherical shell of radius $r$ and half thickness $\Delta r=8\,h_\mathrm{CMB}^{-1}$ Mpc by 
\begin{align}
\Theta_{r}(d(G_i))&=\begin{cases}
1,&\, r-\Delta r\leq d(G_i)<r+\Delta r,\\
0,&\, \hbox{otherwise},
\end{cases}
\end{align}
to smooth away the nonlinear effects. The weighting schemes \cite{Reid:2015gra} for data and random galaxies, respectively,  are assigned as
\begin{align}
w(D_i)&=w_{\mathrm{see},i}w_{\mathrm{star},i}(w_{\mathrm{cp},i}+w_{\mathrm{noz},i}-1)w_{\mathrm{FKP},i},\\
w(R_i)&=w_{\mathrm{FKP},i},
\end{align}
where the Feldman-Kaiser-Peacock (FKP) weight \cite{Feldman:1993ky} $w_{\mathrm{FKP},i}$ is to minimize the theoretical Poisson noises, while the rest of the non-FKP weights \cite{Anderson:2012sa,Anderson:2013zyy} are to minimize the impact of observational artifacts on our estimate of the true galaxy overdensity field. These observational artifacts would affect the completeness of the sample but could be corrected by (i) the total angular systematic weights $w_{\mathrm{systot},i}=w_{\mathrm{star},i}w_{\mathrm{see},i}$ to remove noncosmological fluctuations imprinted on the catalog by the target selection step due to stellar density and seeing, (ii) the weight $w_{\mathrm{cp},i}$ for galaxies not observed due to fibre collisions in a ``close pair'', and (iii) the weight $w_{\mathrm{noz},i}$ for observed galaxies for which a robust redshift was not obtained due to the redshift failure likely to occur on faint targets.

The data and random catalogs \cite{Reid:2015gra,Kitaura:2015uqa} we use to estimate $\delta_g(r)$ come from the final data release (DR12) samples of the Baryon Oscillation Spectroscopic Survey (BOSS) of the Sloan Digital Sky Survey (SDSS) III project \cite{Alam:2015mbd}. Specifically, we use  the pre-reconstructed combined LOWZ and CMASS samples \cite{Alam:2016hwk} with two galactic hemispheres referred as the northern and southern galactic caps (NGC and SGC, respectively). The associated random catalog is created with a number size 50 times larger than the data catalog to minimize the shot noise so that the number counting $N_R(r)$ in calculating $\delta_g(r)$  should be further normalized by a factor of 50. The redshifts for galaxies in the random catalog are assigned by randomly drawing from the measured galaxy redshifts weighted by the total non-FKP weights $w_{\mathrm{tot},i}=w_{\mathrm{systot},i}(w_{\mathrm{cp},i}+w_{\mathrm{noz},i}-1)$. See also \cite{Wang:2017mcf} for an alternative generation of the random catalog with smooth redshift assignments. The evolution of the linear bias for galaxy tracer is borrowed from the detailed reconstructions of matter density fields  \cite{Lavaux:2019fjr} from the same SDSS III BOSS data shown in the inset in the left panel of Fig. \ref{fig:LocalUniverse}. We stress here as a caveat that the bias model used in  \cite{Lavaux:2019fjr} is a nonlocal and nonlinear bias model, while we have used the usual linear bias in Eq. \eqref{eq:bias}. However,  assuming that the dark matter density is mostly the same at the finest and next-to-finest levels of the density fluctuations  \cite{Lavaux:2019fjr}, the nonlinear bias model admits a linear term linking the matter density to the galaxy number count, which is the quantity closest to the usual linear bias. The increasing trend of galaxy bias is expected for magnitude limited surveys since the galaxy-matter bias has been found to increase with luminosity so that the distant galaxies are, on average, more biased than those observed nearby. Therefore, we further adopt a conservative estimation of the bias evolution \cite{Lavaux:2019fjr} with flat extrapolations into the nearby and distant regions. 

\begin{figure*}
\centering
\includegraphics[width=0.9\textwidth]{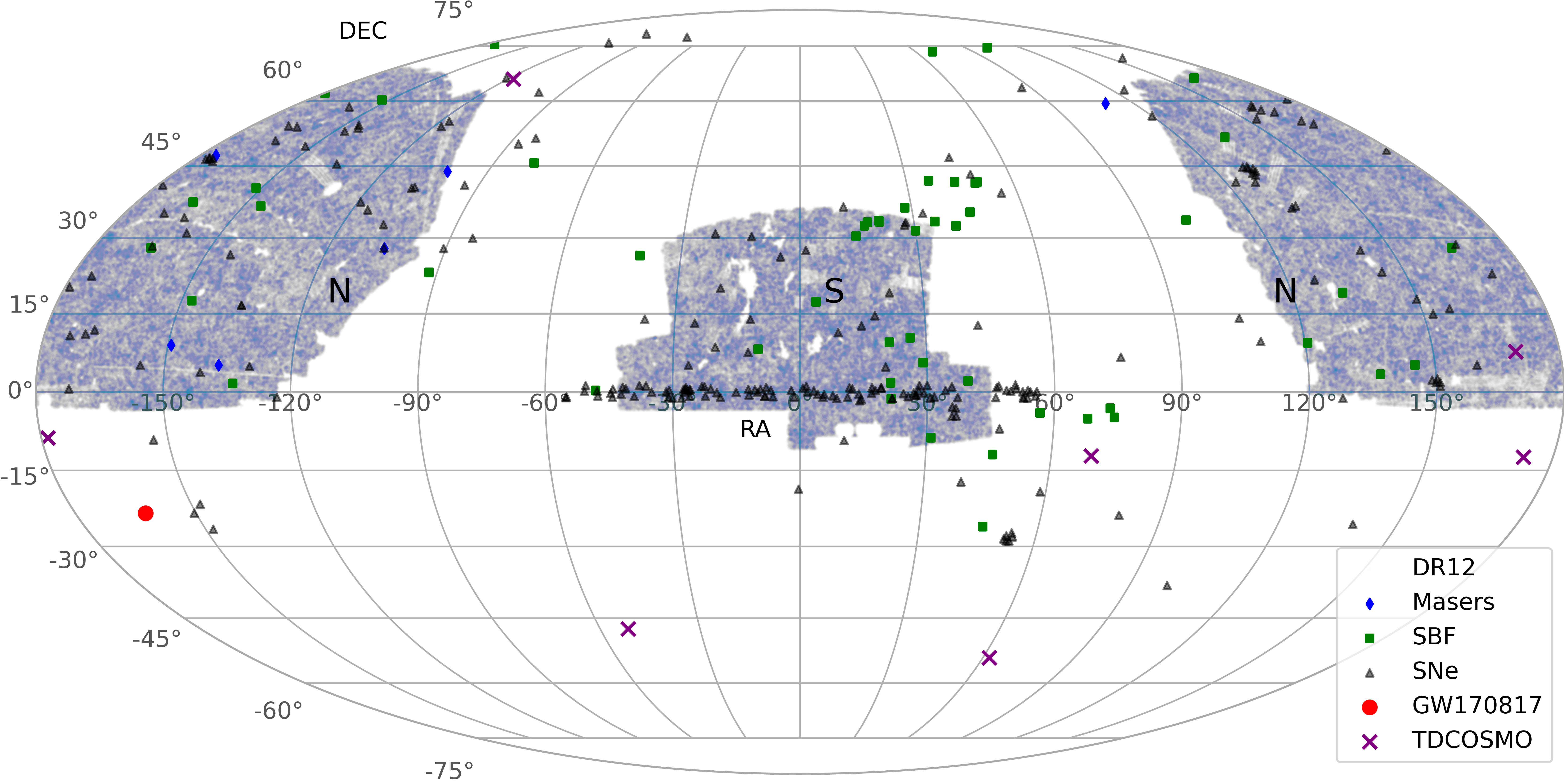}\\
\caption{The Mollweide view of the SNe (black triangles), SBF (green squares), masers (blue diamonds), lensing events (purple crosses) from  the TDCOSMO sample, and the host galaxy NGC4993 of GW170817 (red point) in the equatorial coordinate system. The galaxy survey regions covered by BOSS DR12 are shown as shaded regions, where the central part is in SGC while the rest of the shaded regions are in the NGC.}\label{fig:Sky}
\end{figure*}

The resulted $\Omega_\mathrm{m}(r)$ and the corresponding $H_0(r)$ are shown in the left and right panels of Fig. \ref{fig:LocalUniverse} from the NGC (blue curve) and SGC (red curve) parts of the combined BOSS DR12 LOWZ+CMASS samples, where the bins for plotting are separated at intervals of 0.01 in logarithmic scale. The large fluctuations below  $\sim50$ Mpc might signify the breakdown of LTB isotropy in a sufficiently local Universe ($z\lesssim0.01$), which requires further data analysis from combinations with other near full-sky low-redshift galaxy surveys like the 6dF Galaxy Survey  \cite{Jones:2009yz} and the 2M++ redshift compilation \cite{Lavaux:2011zu,Carrick:2015xza}. Going beyond $\sim50$ Mpc, due to the residual difference in biases between the northern and southern hemispheres, $\Omega_\mathrm{m}(r)$ from the NGC deviates only a few percent from $\Omega_m^\mathrm{CMB}$, shown as the horizontal black dashed line, while $\Omega_\mathrm{m}(r)$ from the SGC admits $10\%-20\%$ deviations around $100\lesssim r\lesssim1000$ Mpc, but still within the $5\sigma$ constraint $|\delta|<27\%$  \cite{Kenworthy:2019qwq} on local density contrasts on scales larger than $69\,h^{-1}$ Mpc set by fitting a simple Garcia-Bellido and Haugb$\o$lle (GBH) profile \cite{GarciaBellido:2008nz} in the $\Lambda$LTB model to combined SNe data from Pantheon sample with the Foundation survey and the most recent release of light curves from the Carnegie Supernova Project. This constraint is likely to be relaxed to allow for $10\%$ of overdensities at the $1\sigma$ level in \cite{Lukovic:2019ryg} using a different GBH profile for the curvature term and different SNe Ia samples from the Joint Light-Curve Analysis  and Pantheon compilation. 

Note that the separate uses of the NGC and SGC data are based on an expectation as if they could separately represent the galaxy number density profile of the whole sky.  However, the suspiciously larger overdensity of SGC than that of NGC might be caused by the lower bias of the SGC than that of the NGC for most of the redshift range of interest. Nevertheless, even without considering any bias, this hemispherical difference is already visible for the pure number counting of data galaxies in the spherical shells of interest from the SGC compared to that of the NGC (after being compensated by a normalization of the total number of data galaxies in each galactic cap), as we have explicitly checked.  We will investigate in a future study this potential hemispherical asymmetry by directly fitting the SNe data to the full time evolution of our model.

The corresponding $H_0(r)$ from the SGC data could incorporate various distance ladder measurements from SNe Ia data calibrated by Cepheid \cite{Riess:2016jrr}, TRGB \cite{Yuan:2019npk}, or Miras \cite{Huang:2019yhh} since there are large overlaps in the space of the SNe data sample with the BOSS SGC region, as shown with black triangles in Fig. \ref{fig:Sky} so that the inferred $\Omega_\mathrm{m}(r)$ could faithfully depict the ambient matter density around those SNe data. It is worth noting that the overall slightly descending trend of $H_0(r)$ out to $\gtrsim1$ Gpc might be related to the similar descending trend of $H_0(r)$ also observed in \cite{Millon:2019slk} for the TDCOSMO-I measurement and in \cite{Krishnan:2020obg} for a flat $\Lambda$CDM model constrained by cosmological data separated by different redshift bins. Nevertheless, the Hubble constant inferred from the far region out to $\gtrsim3$ Gpc should simply be given by the CMB value \cite{Aghanim:2018eyx} with the recovery of the cosmic homogeneity.

However, the implications for other measurements might be inconclusive. The host galaxy NGC4993 for the standard siren measurement of GW170817 \cite{Wang:2020vgr}  is located outside the regions of BOSS DR12, as shown with a red point in Fig. \ref{fig:Sky}. Therefore, the inferred $\Omega_\mathrm{m}(r)$ from the current galaxy survey regions does not correctly account for the local environment of NGC4993. A similar argument could also hold for the SBF samples \cite{Blakeslee:2021rqi}, as shown with green squares in Fig. \ref{fig:Sky}, which are partially located outside the BOSS DR12 regions. However, the situation for the SLTD measurements becomes even more delicate, not only because all the TDCOSMO samples \cite{Birrer:2020tax} except PG1115+080 are located outside regions of BOSS DR12, as shown with purple crosses in Fig. \ref{fig:Sky}, but also each lensing event involves two objects (source and deflector) with a subtlety of choosing which kind of objects as the distance indicator for the TDCOSMO samples. For all the cases mentioned above, the inferred  $\Omega_\mathrm{m}(r)$ might not faithfully reflect the ambient matter density around those datasets. Larger galaxy surveys are needed to make a more convincing comparison with these measurements.

\section{Conclusion and discussions}\label{sec:conclusion}

In conflict with the large-scale structure data, reducing the cosmic sound horizon alone in the early Universe no longer appears to be a compelling resolution of the Hubble tension, and an additional homogeneous dynamical dark energy is strongly constrained by the inverse distance ladder, which, however, could be evaded  by introducing the cosmic inhomogeneities in the Hubble expansion rate at late time from the chameleon coupled to the local matter overdensities. The global-fitting result of the Hubble constant from a CMB measurement is recovered simply because the homogeneity has emerged at cosmic scales. The relatively high values of the Hubble constant from various local measurements could be accommodated by the overdensities observed in galaxy surveys. Larger galaxy surveys are needed to make more accurate predictions on the local Hubble constant profile.

Our result is robust with respect to our test potential of different $n$, and $\alpha$ could be fixed by matching $\Omega_\mathrm{m}^\mathrm{out}$ to the CMB measurement at spatial infinity as expected, leaving the galaxy bias $b$ as the only input free parameter. Our model effectively serves as a zero-parameter model to reconcile both early-time and late-time measurements provided with the more realistic matter density profile measured from other methods even without use of the galaxy bias. 

The current acceleration of our Universe is driven by the quintessential effect of our chameleon dark energy instead of the self-acceleration ruled out by the no-go theorem \cite{Wang:2012kj} with $\Delta\Omega/\Omega\ll1$, as one can explicitly check in our case. Apart from providing an effective cosmological constant, the chameleon field itself also induces a fifth force among matter fields, of which the interaction range estimated by the inverse effective matter $m_\mathrm{eff}^{-1}(r)\equiv V''_\mathrm{eff}(\langle\varphi\rangle(r))^{-1/2}\sim10^{-5}\,\mathrm{m}$ has been numerically verified over the regions of interest, safely meeting the astrophysical/laboratory constraints \cite{Desmond:2018euk,Burrage:2016bwy}. Future experiments \cite{Burrage:2017qrf} are needed for testing our chameleon dark energy model, for example, the ultracold atom experiment in space \cite{Cai:2020lma}.

An overdensity with additional negative pressure introduced by the chameleon dark energy tends to smear at larger scale until reaching a new virial equilibrium due to a decreasing pressure with decreasing matter density. This might shed light on how to  resolve the $S_8$ tension (see, e.g., Ref. \cite{Raveri:2018wln}) by lowering down the matter fluctuation in the vicinity of nonlinear scale. This would require for a full time evolution of our chameleon dark energy at perturbation level, which will be reserved for future work.

\begin{acknowledgments}
We acknowledge the use of data/random catalogs \texttt{galaxy/random0-DR12v5-CMASSLOWZTOT-North/South}, which were downloaded from SDSS BOSS DR12 at website \url{https://data.sdss.org/sas/dr12/boss/lss/}. This work is supported by  National Key Research and Development Program of China Grant No. 2020YFC2201501; National Natural Science Foundation of China Grants No. 11435006, No. 11647601, No. 11690021, No. 11690022, No. 11821505, No. 11851302, No. 12047503, No. 11991052, No. 12075297, and No. 12075298; Strategic Priority Research Program of the Chinese Academy of Sciences (CAS) Grant No. XDB23030100 and No. XDA15020701; the Key Research Program of the CAS Grant No.  XDPB15; and by the Key Research Program of Frontier Sciences of CAS.  
\end{acknowledgments}

\appendix

\section{Chameleon mechanism}\label{app:chameleon}

A chameleon field $\varphi$ coupled to the ambient matter fields $\psi_i$ with conformal factor $\Omega_i(\varphi)$ is defined by the action
\begin{align}
S=&\int\mathrm{d}^4x\sqrt{-g}\left(\frac{M_{\mathrm{Pl}}^2}{2}R-\frac{1}{2}(\partial\varphi)^2-V(\varphi)\right)\nonumber\\
&+\sum\limits_iS^{(i)}_\mathrm{m}\left[\Omega_i^2(\varphi)g_{\mu\nu},\psi_i\right],
\end{align}
of which the equation-of-motion (EoM) of the chameleon field reads
\begin{align}
\nabla^2\varphi&=V'(\varphi)+\sum\limits_iU'_i(\varphi)\equiv V'_\mathrm{eff}(\varphi),
\end{align}
where the original potential $V(\varphi)$ receives an extra contribution of form \cite{Cai:2020ndh}
\begin{align}\label{eq:Ui}
U_i(\varphi)=\Omega_i^{1-3w_i}(\varphi)\hat{\rho}_i.
\end{align}
Here $w_i$ is the equation-of-state (EoS) $p_i=w_i\rho_i$ of the matter field with the energy-momentum tensor $T^\mu_{(i)\nu}=\mathrm{diag}(-\rho_i,p_i,p_i,p_i)$. $\hat{\rho}_i$ is introduced as
\begin{align}
\hat{\rho}_i=\Omega_i^{3w_i-1}\rho_i
\end{align}
in such a way so that it is covariantly conserved in Einstein frame,
\begin{align}
\nabla_t\hat{\rho}_i=0,
\end{align}
which itself is the $\nu=0$ component of the conservation equation,
\begin{align}\label{eq:EinsteinConservation}
\nabla_\mu T^\mu_{(i)\nu}=T_i\Omega_i^{-1}\nabla_\nu\Omega_i,
\end{align}
of the energy-momentum tensor in Einstein frame,
\begin{align}
T_{\mu\nu}^{(i)}\equiv\frac{-2}{\sqrt{-g}}\frac{\delta S_\mathrm{m}^{(i)}}{\delta g^{\mu\nu}}
\end{align}
with $T_i\equiv g_{\mu\nu}T_i^{\mu\nu}$. Note that the energy-momentum tensor defined by
\begin{align}
\tilde{T}_{\mu\nu}^{(i)}&=\frac{-2}{\sqrt{-\tilde{g}_{(i)}}}\frac{\partial}{\partial\tilde{g}^{\mu\nu}_{(i)}}\left(\sqrt{-\tilde{g}_{(i)}}\mathcal{L}_\mathrm{m}^{(i)}[\tilde{g}_{\mu\nu}^{(i)},\psi_i]\right)
\end{align}
is conserved by $\tilde{\nabla}_\mu^{(i)}\tilde{T}^{\mu\nu}_{(i)}=0$ in Jordan frame where $\psi_i$ is minimally coupled to the Jordan-frame metric $\tilde{g}_{\mu\nu}^{(i)}\equiv\Omega_i^2(\varphi)g_{\mu\nu}$. The transformations between these two energy-momentum tensors in two different frames, 
\begin{equation}
\begin{split}
\tilde{T}_{\mu\nu}^{(i)}\Omega_i^2=T_{\mu\nu}^{(i)}, &\quad
\tilde{T}_{(i)}^{\mu\nu}\Omega_i^6=T_{(i)}^{\mu\nu},\\
\tilde{T}^\mu_{(i)\nu}\Omega_i^4=T^\mu_{(i)\nu}, &\quad
\tilde{T}_i\Omega_i^4=T_i, 
\end{split}
\end{equation}
should be straightforward with $\tilde{T}_i\equiv \tilde{g}_{\mu\nu}\tilde{T}_i^{\mu\nu}$. In fact, the conservation equation \eqref{eq:EinsteinConservation} in Einstein frame could be derived from the conservation equation in Jordan frame,
\begin{align}
0=\tilde{\nabla}_\mu^{(i)}\tilde{T}^{\mu\nu}_{(i)}=\Omega_i^{-6}\nabla_\mu T^{\mu\nu}_{(i)}-T_i\Omega_i^{-7}\nabla^\nu\Omega_i,
\end{align}
where we have used 
\begin{align}
C^{\rho(i)}_{\mu\nu}=\Omega_i^{-1}(\delta^\rho_\mu\nabla_\nu\Omega_i+\delta^\rho_\nu\nabla_\mu\Omega_i-g_{\mu\nu}g^{\rho\lambda}\nabla_\lambda\Omega_i)
\end{align}
for the connection $\tilde{\Gamma}^{\rho(i)}_{\mu\nu}=\Gamma^\rho_{\mu\nu}+C^{\rho(i)}_{\mu\nu}$ in Jordan frame.

\section{LTB Universe with dynamical inhomogeneity}\label{app:LTB}

We approximate our local Universe with an inhomogeneous but isotropic Lema\^{i}tre-Tolman-Bondi (LTB) metric \cite{Lemaitre:1933gd,Tolman:1934za,Bondi:1947fta} 
\begin{align}
\mathrm{d}s^2=-\mathrm{d}t^2+\frac{A'(t,r)^2}{1-k(r)}\mathrm{d}r^2+A(t,r)^2\mathrm{d}\Omega_2^2
\end{align}
with $A'(t,r)=\partial A(t,r)/\partial r$, where the usual Friedmann-Lema\^{i}tre-Robertson-Walker (FLRW) metric is recovered for $A(t,r)=a(t)r$ and $k(r)=kr^2$ with $a(t)$ the usual scale factor. The $(t,t)$- and $(r,r)$-components of the Einstein's equations in the LTB metric read
\begin{align}
\frac{\dot{A}^2+k(r)}{A^2}+\frac{2\dot{A}\dot{A}'+k'(r)}{AA'}&=8\pi G(\hat{\rho}_\mathrm{m}+\hat{\rho}_\Lambda),\label{eq:FRW1}\\
\dot{A}^2+2A\ddot{A}+k(r)&=8\pi G\hat{\rho}_\Lambda A^2,\label{eq:FRW2}
\end{align}
respectively, the second of which, after multiplied by $\dot{A}\equiv\partial A(t,r)/\partial t$,
could be directly integrated as
\begin{align}\label{eq:FRW12}
\frac{\dot{A}^2}{A^2}+\frac{k(r)}{A^2}=\frac{8\pi G}{3}\hat{\rho}_\Lambda+\frac{F(r)}{A^3}-\frac{8\pi G}{3A^3}\int\mathrm{d}t(A^3\partial_t\hat{\rho}_\Lambda).
\end{align}
Here $F(r)$ is an integration ``constant'' of the time derivative determined by matching the spatial derivative of \eqref{eq:FRW12} to \eqref{eq:FRW1} as
\begin{align}\label{eq:Fprime}
3M_\mathrm{Pl}^2F'(r)=\hat{\rho}_\mathrm{m}\partial_r A^3-A^3\partial_r\hat{\rho}_\Lambda+\int\mathrm{d}t\,\partial_r(A^3\partial_t\hat{\rho}_\Lambda).
\end{align}
Combining \eqref{eq:FRW1}-\eqref{eq:Fprime} gives rise to the modified equation for the $(r,r)$-component of the Einstein's equations
\begin{align}\label{eq:FRW22}
\frac23\frac{\ddot{A}}{A}+\frac13\frac{\ddot{A}'}{A'}=\frac{8\pi G}{3}\hat{\rho}_\Lambda&-\frac{4\pi G}{3}\left(\hat{\rho}_\mathrm{m}-\frac{3A^3\partial_r\hat{\rho}_\Lambda}{\partial_rA^3}\right).
\end{align}
Therefore, the LTB version of the first Friedmann equation \eqref{eq:FRW12} depends on the dynamical property of chameleon dark energy, while the LTB version of the second Friedmann equation \eqref{eq:FRW22} depends on its inhomogeneity.

Eq.  \eqref{eq:FRW12} could be cast into a Friedmann-like form
\begin{align}\label{eq:FRW}
\frac{H^2}{H_\infty^2}=\Omega_\mathrm{m}\left(\frac{A_0}{A}\right)^3+\Omega_\Lambda+\Omega_k\left(\frac{A_0}{A}\right)^2
\end{align}
with $A_0(r)\equiv A(t_0,r)$, $H(t,r)=\dot{A}(t,r)/A(t,r)$, $H_\infty\equiv H(t_0,r\to\infty)\equiv H_\mathrm{CMB}$, and
\begin{align}
\Omega_k(r)=-\frac{k(r)}{A_0^2(r)H_\infty^2}, &\quad \Omega_\Lambda(t,r)=\frac{\hat{\rho}_\Lambda(t,r)}{3M_\mathrm{Pl}^2H_\infty^2},
\end{align}
\begin{align}\label{eq:Omegam}
\Omega_\mathrm{m}(t,r)=\frac{1}{A_0^3(r)H_\infty^2}\left(F(r)-\frac{8\pi G}{3}\int\mathrm{d}t\,(A^3\partial_t\hat{\rho}_\Lambda)\right).
\end{align}
Note that the relation between $\Omega_\mathrm{m}(t,r)$ and $\hat{\rho}_\mathrm{m}(t,r)$ is constrained by \eqref{eq:Fprime}, namely,
\begin{align}\label{eq:Omrm}
3M_\mathrm{Pl}^2H_\infty^2\partial_r(\Omega_\mathrm{m}A_0^3)=\partial_rA^3\left(\hat{\rho}_\mathrm{m}-\frac{A^3\partial_r\hat{\rho}_\Lambda}{\partial_r A^3}\right),
\end{align}
which reduces to the usual definition of the energy density fraction $\Omega_\mathrm{m}=\hat{\rho}_\mathrm{m}/(3M_\mathrm{Pl}^2H_\infty^2)$ for homogeneous components in FLRW Universe with $A(t,r)=a(t)r$. However, for general dynamical inhomogeneous setting, there is no \textit{a priori} simple relation between $\Omega_\mathrm{m}(t,r)$ and $\hat{\rho}_\mathrm{m}(t,r)$ without first solving $A(t,r)$ from \eqref{eq:FRW}, which itself contains both $\Omega_\mathrm{m}$ and $\hat{\rho}_\mathrm{m}$ coupled to chameleon in $\Omega_\Lambda$. Nevertheless, \eqref{eq:Fprime} could be directly integrated by appreciating the definition \eqref{eq:Omegam} as
\begin{align}
3M_\mathrm{Pl}^2H_\infty^2A_0^3\Omega_\mathrm{m}=\hat{\rho}_\mathrm{m}A^3-\int\mathrm{d}r\,A^3\partial_r(\hat{\rho}_\mathrm{m}+\hat{\rho}_\Lambda),
\end{align}
which, under a condition of a slowing varying dark energy in radial profile compared to the matter density,
\begin{align}\label{eq:flatDE}
\frac{1}{3\hat{\rho}_\mathrm{m}}\frac{\partial_r\hat{\rho}_{\Lambda}}{\partial_r\ln A}\ll1,
\end{align}
becomes
\begin{align}
3M_\mathrm{Pl}^2H_\infty^2A_0^3\Omega_\mathrm{m}=\hat{\rho}_\mathrm{m}A^3.
\end{align}
Therefore,
\begin{align}
\Omega_\mathrm{m0}(r)\equiv\Omega_\mathrm{m}(t_0,r)=\frac{\hat{\rho}_\mathrm{m}(t_0,r)}{3M_\mathrm{Pl}^2H_\infty^2}\equiv\frac{\hat{\rho}_\mathrm{m0}(r)}{3M_\mathrm{Pl}^2H_\infty^2}
\end{align}
after evaluated at the present time $t_0$. We have verified the condition \eqref{eq:flatDE} at $t_0$  numerically in the region of interest for the realistic matter density profile of our local Universe. 

\bibliographystyle{utphys}
\bibliography{ref}

\end{document}